\RequirePackage{fix-cm}
\documentclass[smallextended]{svjour3}       %
\smartqed  % flush right qed marks, e.g. at end of proof
\usepackage{graphicx}
\usepackage{xcolor}
\usepackage{amsmath}
\usepackage{url}

\journalname{DRAFT}
\begin{document}

\title{Session-Based Hotel Recommendations:\\Challenges and Future Directions}
% Towards Hotel Recommendation in Booking Sessions: Challenges and Future Directions
\subtitle{As part of the ACM Recommender System Challenge 2019}

\titlerunning{Session-Based Hotel Recommendations}        % if too long for running head

\author{Jens Adamczak        \and
        Gerard-Paul Leyson   \and
        Peter Knees          \and 
        Yashar Deldjoo       \and
        Farshad Bakhshandegan Moghaddam \and
        Julia Neidhardt \and 
        Wolfgang W\"orndl \and
        Philipp Monreal}


\institute{J. Adamczak \at
           trivago N.V., D\"usseldorf, Germany,  \email{jens.adamczak@trivago.com}
           \and
           G. Leyson \at
           trivago N.V., D\"usseldorf, Germany, 
           \email{gerard-paul.leyson@trivago.com}
           \and
           P. Knees \at
           %Faculty of Informatics, 
           TU Wien, Austria, 
           \email{peter.knees@tuwien.ac.at}           %  \\
%          \emph{Present address:} of F. Author  %  if needed
           \and
           Y. Deldjoo \at
           Polytechnic University of Bari, Italy, 
           \email{yashar.deldjoo@poliba.it}           %  \\
           \and
           F. Bakhshandegan Moghaddam \at
           Karlsruhe Institute of Technology, Germany, 
           \email{farshad.moghaddam@student.kit.edu}
           \and
           J. Neidhardt \at
           %Faculty of Informatics, 
           TU Wien, Austria, 
           \email{julia.neidhardt@ec.tuwien.ac.at}           %  \\
           \and
           W. W\"{o}rndl \at
           Technical University of Munich, Germany, 
           \email{woerndl@in.tum.de}           %  \\
           \and
           P. Monreal \at
           trivago N.V., D\"usseldorf, Germany, 
           \email{philipp.monreal@trivago.com}
}

\date{Received: date / Accepted: date}

\maketitle

\begin{abstract}
%\yashar{Please improve/extent the abstract if needed; also it looks similar to RecSys Challenge 2018 overview paper.}\peter{gave it a (first) reframing.}
In the year 2019, the Recommender Systems Challenge deals with a real-world task from the area of e-tourism for the first time, namely the recommendation of hotels in booking sessions. 
In this context, this article aims at identifying and investigating what we believe are important domain-specific challenges recommendation systems research in hotel search is facing, from both academic and industry perspectives. 
We focus on three main challenges, namely dealing with \emph{(1)} multiple stakeholders and value-awareness in recommendations,
\emph{(2)} sparsity of user data and the extensive cold-start problem, and \emph{(3)} dynamic input data and computational requirements.
To this end, we review the state of the art toward solving these challenges and discuss shortcomings. 
We detail possible future directions and visions we contemplate for the further evolution of the field. 
This article should, therefore, serve two purposes: giving the interested reader an overview of current challenges in the field and inspiring new approaches for the ACM Recommender Systems Challenge 2019 and beyond. 
%\yashar{Till here}
\end{abstract}

\section{Introduction}
\label{intro}

% Tourism recommendation, state of the art, particularities (complex, multi-faceted), linking to RSC’19 tasks/challenges 

The rapid development of information and communication technologies and the web have transformed the tourism and travel domain. 
Today, travelers no longer rely on travel agencies but search for information themselves and compose their trips according to their specific preferences. 
Users have to choose from a multitude of options and recommender systems for travel and tourism can be practical tools to overcome the inevitable information overload. 
Such developments in e-tourism have been studied at the RecSys Conference\footnote{\url{https://recsys.acm.org}} and the RecSys Workshop on Recommenders in Tourism (RecTour)\footnote{\url{http://www.ec.tuwien.ac.at/rectour2019/}}  series, for example. 
The travel industry has moved from separate websites to IT-based business networks and a ``customer-driven'' field looking at online communities or social media sites and their respective power \cite{DBLP:journals/jitt/NeidhardtW18}.

Recommending hotels and other travel-related items is still a difficult task as travel and tourism is a very complex domain. 
Planning a trip usually involves searching for a set or package of products that are interconnected (e.g., means of transportation, lodging, attractions), with rather limited availability, and where contextual aspects may have a major impact (e.g., time, location, social context). 
Users book much fewer hotels than, say, listen to music tracks and, given the financial obligation of booking a stay at a hotel, users usually exhibit a strong price sensitivity and a bigger need to be convinced by any given offer. 
Besides, travelers are often emotionally connected to the products and the experience they provide. 
Therefore, decision making is not only based on rational and objective criteria. 
As such, providing the right information to visitors of a travel site, such as a hotel booking service at the right time about their items and various other services, is challenging.

Information about items such as hotels is often available as item metadata.
Text analysis and opinion mining techniques can be used to extract additional insights about items from reviews, such as discussed topics and opinions, contextual information and reviewers’ emotions \cite{DBLP:journals/umuai/ChenCW15}. 
However usually, in this domain, information about users and their goals and preferences are harder to obtain. 
Systems need to analyze session-based data of anonymous users to tailor information search and predict the hotels the users may be interested in. 

Recommendation systems for hotels and accommodations exist in different forms and scopes of application. 
In the following, we will focus on three challenges pertinent to the scenario of search result ranking for user sessions as presented in the RecSys Challenge 2019\footnote{\url{http://www.recsyschallenge.com/2019/}}.
In the given scenario, the goal is to develop a session-based and context-aware recommender system using various input data to provide a list of accommodations that will match the needs of the user.
Participants of the challenge are tasked with predicting which accommodations (items) have been clicked in the search result during the last part of a user session.
For building and testing models, a public dataset of real-world hotel search sessions has been released by trivago.\footnote{\url{https://www.trivago.com}}

%\yashar{The remainder of the article is structured as following:}
%... [now lead over to specific RSC’19 task] ...

\section{Challenges}
\label{challenges}
Recommender systems for the hotel search domain have to address the issues mentioned above by considering theoretical and practical aspects of providing adequate suggestions to users. Effective algorithms, useful objective functions and metrics that capture the essence of complex user search patterns and business contexts need to be developed. In the following, we will present three specific challenges that demonstrate the particularities of providing hotel recommendations to users in a session-based setting.

\subsection{\textbf{Challenge 1: Multiple stakeholders and need for value-aware recommendations}}\label{sec:challenge1}

\subsubsection{Problem definition}
Many modern online services in the online travel domain such as hotel booking sites, Online Travel Agencies (OTAs), and price comparison and metasearch sites, represent two-sided marketplaces that have to balance supply and demand for accommodations. Recommender systems developed in this domain are tasked with considering multiple objectives to satisfy user needs, supplier needs and ensure effective monetization of the service. Since most of these sites act as intermediaries between the user and the suppliers of the offered product - the hotels themselves - they have to further take into consideration the interests of multiple stakeholders to create an environment with high incentives for all participants \cite{Jannach:2016:RP:2959100.2959186,Krasnodebski:2016:CSR:2959100.2959124}. For metasearch and price comparison services, this is of special importance since the suppliers usually compete with each other in an auction to forward traffic to their sites and run profitable campaigns.

\subsubsection{State-of-the-art}
The theoretical research on value-aware recommendation systems is in its early stages. The workshop on value-aware and multi-stakeholder recommendation at the RecSys conference 2017 provides a good overview of current challenges in the area \cite{Burke2017VAMS2W}. Apart from attempts to characterize the problem and to develop a taxonomy for the field, the workshop also highlighted practical approaches, such as a learning-to-rank approach aimed at finding an optimal weighting between different components of a multi-stakeholder objective function \cite{Nguyen2017AML}.

\subsubsection{Limitations}
One of the difficulties in addressing multiple stakeholders is the need for suitable metrics that adequately capture the reaction of the marketplace participants to changes applied to the recommender. Good metrics would promote recommendations where most or, ideally, all stakeholders are satisfied. The task of finding suitable metrics is complicated by the difficulty of properly evaluating the short- and long-term effects on the marketplace. For example, changes in bidding behavior of suppliers and elasticity of the market are difficult to measure offline or in typical A/B-test scenarios and often require special attention.

\subsection{\textbf{Challenge 2: Sparsity of user data and extensive cold-start problem}}\label{sec:challenge2}

\subsubsection{Problem definition}
Due to the nature of the tourism domain, hotel recommender systems have to be able to deal with unreliable and sparse input data. The reasons are varied. 
Users usually do not frequently go on vacation and can have long intervals between bookings. Returning users of a recommender service can have changed preferences or unstable preferences to begin with, e.g. when using the service for both business and leisure travel. 
Due to the high price of transportation and lodging, users compare offers from different services in extensive user journeys across multiple websites that are only partially accessible for any given service. Metasearch services, in particular, have particularly poor visibility on user information since they often do not require the user to make an account or provide additional user data that can be used to feed into their recommendation systems. Since the booking does not take place on their site, they have to rely on additional tracking data that has to be provided by the actual booking site. Furthermore, a subset of users still prefers to call hotels directly to make a reservation, which is very hard to track.

\subsubsection{State-of-the-art}
Cold-start is a well-studied problem in recommender system research, but established solutions mostly focus on providing initial recommendations for new users or items and reach their limits when overall little user data is available and most of the users can be considered new for extended periods of time. It is crucial from an industry perspective to reduce the cold-start problem through data collection and to build recommendation systems that are targeted to the limited available information of a user at runtime. To collect more data, additional ways have to be revised to effectively identify users across their journey on different websites and products \cite{DBLP:journals/corr/abs-1710-08217}. From the algorithmic perspective, recommender systems that are based on the explicit and implicit signals that users leave behind in the sequential interaction patterns in a session will be particularly useful \cite{DBLP:journals/corr/abs-1803-09587}. Sparsity can be addressed with Bayesian methods that provide reasonable default values for long-tail items and help to quantify the uncertainty in noisy environments.

\subsubsection{Limitations}
Session-based recommendations per definition rely heavily on the interaction of users with exposed items. The quality of the recommendations will therefore suffer if no matching items are displayed to the user to begin with. This puts strong constraints on the initial selection of items and requires a good balance between exploration and exploitation and mechanisms that generate a pool of suitable items supporting the session-based recommender. To deal with a large number of listings that could potentially be shown to the user, computationally effective candidate selection methods can be employed that use information retrieval techniques to detect the intent of a search query~\cite{Arya:2017:CSL:3077136.3082066}.

\subsection{\textbf{Challenge 3: Dynamic input data and computational requirements}}\label{sec:challenge3}

\subsubsection{Problem definition}
The online travel market is highly competitive and dynamic. For example, travel agencies are looking to attract users by advertising their hotel rooms on multiple performance marketing channels and let users do their booking on a first come, first served basis. Hotels have dynamic pricing models aimed at optimizing their profit. Hotel rooms are perishable goods; hotels have rather fixed costs whether the rooms are filled or not. This creates strong incentives to adapt prices frequently, especially for short time-to-arrival. In this scenario, successful recommender systems need to incorporate up-to-date information and serve them with low response times to the users to avoid the frustration of encountering unavailable offers or unexpected prices.

\subsubsection{State-of-the-art}
This set of problems is very specific to industry related applications and has therefore not gotten a lot of attention in academic research. Implementations of recommendation systems in commercial settings depend on the availability of fast infrastructure and intelligent caching systems. Furthermore, the task to provide fast and suitable recommendations is often separated into extensive offline feature creation and model training, and a subsequent live calculation of predictions to maximize the business metrics of interest such as clicks or bookings \cite{Grbovic:2018:RPU:3219819.3219885}.

\subsubsection{Limitations}
Limitations are set by the complexity of the underlying models to predict the user response. Techniques described in the literature, such as latent feature vector approaches based on embedding techniques or matrix factorization, are difficult to employ when item characteristics are prone to frequent changes in the live context.

\section{Performance Indicators}
\label{performance}
%\ac{some references would be good to be added here}

The variety of challenges makes it hard to turn to universally applicable metrics and performance indicators to judge the quality of the developed solutions. Conventional A/B testing frameworks~\cite{DBLP:conf/wsdm/GilotteCNAD18} 
%\wolfgang{cite Alexandre Gilotte, Clément Calauzènes, Thomas Nedelec, Alexandre Abraham, and Simon Dollé. 2018. Offline A/B Testing for Recommender Systems. In Proceedings of the Eleventh ACM International Conference on Web Search and Data Mining (WSDM '18). ACM, New York, NY, USA, 198-206. DOI: https://doi.org/10.1145/3159652.3159687} 
are often used to evaluate how well the recommendations meet the business requirements, but come with limitations and risks, e.g. time and sample size constraints and a potentially less than optimal user experience for tests that are in the early stages of development. To increase the chance to expose only promising tests to the users and to maximize the speed of iteration, algorithms are typically first evaluated offline. For both scenarios, metrics can be used that allow inspecting different aspects of the recommendation problem \cite{Agarwal:2016:SMR:3019548}.

\subsection{Individual prediction metrics} 
Some of the metrics aim at quantifying how well a predicted value matches an observed quantity. In classical recommender settings, this quantity is often the rating a user can assign to a particular item~\cite{DBLP:reference/rsh/2011}. %\wolfgang{cite Shani G., Gunawardana A. (2011) Evaluating Recommendation Systems. In: Ricci F., Rokach L., Shapira B., Kantor P. (eds) Recommender Systems Handbook. Springer, Boston, MA}. 
In the case of hotel recommendations as presented in the current context, the observation is a user interaction with a recommended accommodation (item) of any kind that has to be predicted, e.g. the consumption of content, the click out to an advertiser website, or the booking of an accommodation. \\
 
\textbf{Root mean square error (RMSE)}: The root mean square error measures the difference between the predicted value and the observed one, giving larger weight to larger deviations~\cite{DBLP:reference/rsh/2011}.
%\wolfgang{cite Shani G., Gunawardana A. (2011) Evaluating Recommendation Systems. In: Ricci F., Rokach L., Shapira B., Kantor P. (eds) Recommender Systems Handbook. Springer, Boston, MA}. 
The RMSE is typically used in regression scenarios with numerical outcomes, e.g. to compare a predicted aggregated number of clicks or bookings for a given accommodation to the observed one over a predefined time frame. The metric is also useful to measure the quality of predictions for binary outcomes. For example, let $\hat{y}_i$ be the predicted probability that an impressed item $i$ will result in a click or booking and $y_i$ the observed outcome, i.e. click/booking or no click/no booking. Then the RSME for a set of $N$ impressions can be calculated as

\begin{align}
RMSE = \sqrt{\frac{\sum \limits_{i=1}^{N} \left(\hat{y}_i - y_i \right)^2 }{N}}.
\end{align}

\textbf{Cross entropy loss}: The cross entropy loss or log loss is a fitting choice in settings in which the predicted value is a probability between 0 and 1 and the observed value is a binary outcome that has to be classified~\cite{DBLP:journals/csur/ZhangYST19}.
%\wolfgang{cite Shuai Zhang, Lina Yao, Aixin Sun, and Yi Tay. 2019. Deep Learning Based Recommender System: A Survey and New Perspectives. ACM Comput. Surv. 52, 1, Article 5 (February 2019), 38 pages. DOI: https://doi.org/10.1145/3285029}. 
Log loss penalizes an incorrect classification but does so much more heavily for incorrect classification when the predicted signal was strong. In the notation above log loss can be defined as

\begin{align}
LogLoss = -y_i \log (\hat{y}_i) - (1 - y_i)\log(1 - \hat{y}_i)
\end{align}

\textbf{Area under the ROC curve (AUC)}: The AUC is usually defined in terms of the Receiver Operating Characteristic Curve (ROC) which is a two-dimensional curve that is generated by plotting the false positive rate vs. the true positive rate of a binary classification for varying thresholds of the classifier ~\cite{BAMBER1975387}.
%\wolfgang{cite Bamber, D.: The area above the ordinal dominance graph and the area below the receiver operating characteristic graph. Journal of Mathematical Psychology 12, 387–415 (1975)}. 
In a probabilistic interpretation, the AUC measures the chance that a random positive event (e.g. an item was clicked, an accommodation was booked) was assigned a higher predicted probability than a random negative event (i.e. no click or no booking). In an ideal case, the predicted probability for events that turned out to be true should always be higher than the probabilities predicted for events that turned out to be false, i.e. a higher AUC is preferred. If an algorithm returns $N_+$ samples in the positive class and $N_-$ samples in the negative class and results are ordered according to the algorithm's predictions, the AUC can be defined in Iverson notation as

\begin{align}
AUC = \frac{1}{N_+ N_-} \sum \limits_{k < i} \left[ y_{(k)} < y_{(i)} \right],
\end{align}

with the observed outcomes $y$.

As compared to metrics like RMSE and log loss that measure the absolute deviation of a predicted value from an observed once, AUC is measuring the quality of the relative ranking of the predicted events.

\subsection{Ranking metrics}
The metrics mentioned above allow one to draw conclusions about how well a given model can predict individual events. In many cases, predictive models are used to deliver a final output list of recommendations that can be presented to the users. For these scenarios, additional metrics are necessary that consider the format of the presentation~\cite{DBLP:reference/rsh/2011}. 
%\wolfgang{cite Shani G., Gunawardana A. (2011) Evaluating Recommendation Systems. In: Ricci F., Rokach L., Shapira B., Kantor P. (eds) Recommender Systems Handbook. Springer, Boston, MA}. \\
 
\textbf{Mean reciprocal rank (MRR)}: For a list of ranked items, the reciprocal rank is the multiplicative inverse of the rank of the first positive response. For example, if an item $i$ is clicked or booked on position $\text{rank}_i$ in the result list (counting from the top), the reciprocal rank is denoted as 1/$\text{rank}_i$. The mean reciprocal rank is the average of all reciprocal ranks for a given number $N$ of inspected results lists,

\begin{align}
MRR = \frac{1}{N} \sum \limits_{i=1}^{N} \frac{1}{\text{rank}_i}
\end{align}

\textbf{Normalized discounted cumulative gain (NDCG)}: The MRR considers only the position of the first relevant item in a list. In practice, often situations arise, in which multiple relevant items appear in a list or different levels of relevance should be evaluated, e.g. content consumption, click out to an advertiser page, or the booking of an item. In these cases, the discounted cumulative gain (DCG) penalizes result lists in which relevant items appear at lower positions $p$ in the list~\cite{Jarvelin:2002:CGE:582415.582418}.
%\wolfgang{cite Järvelin, K., Kekäläinen, J.: Cumulated gain-based evaluation of ir techniques. ACM Trans. Inf. Syst. 20(4), 422–446 (2002). DOI http://doi.acm.org/10.1145/582415.582418}. 
For example, the DCG at position $p$ in the list can be calculated as

\begin{align}
DCP_p = \sum \limits_{i=1}^{p} \frac{2^{rel_i} - 1}{\log_2 (i + 1)},
\end{align}

where $rel_i$ is the relevance assigned to the event at position $i$. To evaluate the quality of multiple result lists that can vary in length in practice often the DCG is often normalized by the ideal DCG that can be achieved on a particular position.

\begin{align}
nDCG_p = \frac{DCP_p}{IDCG_p},
\end{align}

where $IDCG_p$ is the DCG of the list ordered by the relevance of the items up to position $p$.

\subsection{List comparison metrics}
All metrics described so far help to understand the quality of an individual prediction or result list presented to the user. To diagnose the composition of result lists and to better understand the effect of an algorithm change on the final result, list comparison metrics can be useful.\\

\textbf{Spearman's rank correlation coefficient}: The rank correlation coefficient according to Spearman $\rho$ measures the correlation between two ranked variables~\cite{HERNANDEZDELOLMO2008790}.
%\wolfgang{cite F. Hernández, E. Gaudioso, Evaluation of recommender systems: a new approach, Expert Systems with Applications (35) (2008) 790–804, doi:10.1016/j.eswa.2007.07.047}. 
In the example of two lists of $N$ ranked accommodations, the correlation of the two lists can be determined via the difference in rank $d_i$ for each item $i$ that appears in both lists

\begin{align}
\rho = 1 - \frac{6}{N (N^2 - 1)} \sum \limits_{i=1}^{N} d_i
\end{align}

\textbf{Jaccard similarity}: Spearman's rank correlation is useful to compare the agreement of two ranked lists, but relies on the assumption that both lists contain the same set of items. However, in cases where only a subset of items is considered, e.g. items on top positions of a rank list, the sets of items can be compared to determine how many items appear in both lists. The Jaccard similarity $J$ of two lists $L1$ and $L2$ is defined by the ratio of the intersection and the union of the elements in the lists~\cite{DBLP:conf/sigmod/KoutrikaBG09}.
%\wolfgang{cite Georgia Koutrika, Benjamin Bercovitz, and Hector Garcia-Molina. 2009. FlexRecs: expressing and combining flexible recommendations. In Proceedings of the 2009 ACM SIGMOD International Conference on Management of data (SIGMOD '09), Carsten Binnig and Benoit Dageville (Eds.). ACM, New York, NY, USA, 745-758. DOI: https://doi.org/10.1145/1559845.1559923}.

\begin{align}
J(L1, L2) = \frac{|L1 \cap L2|}{|L1 \cup L2|}
\end{align}

\section{Future directions}
\label{vision}
%\philipp{I replaced hotel by accommodation in the following paragraph}
There is a multitude of interesting problems to explore further that are only addressed briefly in this article. Work on session-based recommendation algorithms has already been conducted in industry and academia. The applications vary widely and touch different areas from e-commerce to the tourism domain \cite{DBLP:journals/corr/abs-1802-08452}. Accommodation recommendations can profit from this field of research if further algorithm improvements are targeted to the specifics of the user search patterns that can typically be identified on travel websites. The data that describes the characteristics of accommodation inventory and user interaction types that are meaningful in this context need to be made available for researchers and practitioners and requires strong collaboration between industry and academia. As a general theme, the consideration and integration of relevant business metrics in the development and evaluation of algorithmic approaches for accommodation recommendations could benefit from a stronger focus and would find a direct field of application in many tourism-related domains. Value-aware recommendations are a promising step in this direction and can have direct business implications. Furthermore, most approaches try to accurately predict the next item a user is likely interested in, but additional issues such as coverage, diversity and serendipity of recommendations have to be taken into account as well~\cite{DBLP:conf/recsys/GeDJ10}.
%\wolfgang{cite Mouzhi Ge, Carla Delgado-Battenfeld, and Dietmar Jannach. 2010. Beyond accuracy: evaluating recommender systems by coverage and serendipity. In Proceedings of the fourth ACM conference on Recommender systems (RecSys '10). ACM, New York, NY, USA, 257-260. DOI: https://doi.org/10.1145/1864708.1864761}. \ac{feel free to enrich with some useful citations}

\begin{acknowledgements}
We would like to thank all researchers in the fields of recommender systems and tourism and travel recommender systems, with whom we had the pleasure to discuss and collaborate to write this article. We would like to thank everyone at Trivago who was additionally involved in the RecSys Challenge 2019, including Matthias Endler, Simon Br\"uggen, and Wolfgang Gassler.  
Furthermore, we greatly appreciate the help of D\'{a}vid Zibriczky and the previous organizers of the RecSys Challenge, in particular by Markus Schedl and Hamed Zamani. 
\end{acknowledgements}

\bibliographystyle{ieeetr}
\bibliography{refs}
\end{document}